\documentclass[12pt,amsfonts]{article}  
\usepackage{amsfonts}
\usepackage{amsmath}
\def\ts{\textstyle}
\def\t{\textstyle}        

\def\half{{\textstyle{\frac{1}{2}}}}

\def\quarter{\textstyle{\frac{1}{4}}}

\def\H{{\cal H}}

\def\p{\phi}

\def\H{{\cal H}}

\def\l{\lambda}

\def\S{\Sigma'}

\def\k{\kappa}

\def\tr{\rm Tr}

\def\ra{\rightarrow}

\def\tint{{\textstyle\int}}

\def\s{\hskip.08em}
\def\d{\partial}

\def\a{\alpha}
\def\b{\begin{eqnarray}}  
\def\e{\end{eqnarray}}    
\def\bn{\begin{eqnarray}}  
\def\en{\end{eqnarray}}   
\def\<{\langle}
\def\>{\rangle}
\def\dn{d^n\!x}
\def\no{\nonumber}

\def\k{\kappa}

\def\{{\lbrace}
\def\}{\rbrace}
\title{Nontrivial Quantization of $\phi^4_n$, $n\ge2$}
\author{John R. Klauder\footnote{Email: klauder@phys.ufl.edu}\\
Department of Physics and\\Department of Mathematics\\
University of Florida\\
Gainesville, FL 32611-8440}
\date{ }
\begin{document}
\maketitle
\begin{abstract}
Conventional quantization of covariant scalar field models $\p^4_n$, for spacetime dimensions $n\ge5$ is trivial,
 and this may also be true for $n=4$ as well. However, an alternative ${\cal O}(\hbar)$ counterterm leads to nontrivial results for all $n\ge4$, and provides a different quantization for $n=2,3$ as well. In this article we determine the counterterm that provides these desirable properties as simply and directly as possible. The same counterterm also resolves models such as $\p^p_n$ for all even $p$, including those where $p>2n/(n-2)$, which are traditionally regarded as nonrenormalizable.
\end{abstract}
\subsection*{Introduction}
Conventional canonical quantization of the scalar field models $\p^4_n$ is trivial for spacetime dimensions $n\ge5$ \cite{aiz,fro}, possibly trivial for $n=4$ \cite{wei,wil}, and nontrivial for $n=2,3$ \cite{gj}. More generally, the models $\p^p_n$, for $p>2n/(n-2)$ are nonrenormalizable and ill-defined by a conventional perturbation analysis \cite{qft}. Conventional quantization is ambiguous up to ${\cal O}(\hbar)$ terms, which are normally `decided by experiment';  
for example, an harmonic oscillator normally has a positive zero-point energy except when it is one of the degrees of freedom of a free field. Our goal is to find an {\it unconventional} ${\cal O}(\hbar)$ counterterm for scalar field models, and, specifically, we will outline the nontrivial quantization of $\p^4_n$ for all $n\ge2$, including an alternative quantization for $n=2,3$.

The family of models we consider has 
a classical (but imaginary time) action functional
\bn I(\p)=\tint\{\half[(\nabla\p)(x)^2+m_0^2\s\p(x)^2]+g_0\s\p(x)^4\s\}\,d^n\!x\;. \label{e1}\en
In turn, the quantization of this field may be addressed by the formal functional integral
  \bn S(h)\equiv{\cal M}\int e^{\t (1/\hbar)\tint h(x)\s\p(x)\,d^n\!x-(1/\hbar)[I(\p)+ F(\p,\hbar)/2]}
  \,\Pi_x \s d\p(x)\;,\label{e2}\en
  where ${\cal M}$ is chosen so that $S(0)=1$, $h(x)$ is a source function, and $F(\p,\hbar)$ represents the as-yet-unspecified counterterm to control divergences, which should formally vanish as $\hbar\ra0$ so that the proper classical limit formally emerges. Traditionally, a perturbation expansion of the quartic interaction term guides the choice of the counterterm in order to cancel any divergences that are encountered. Here, on the other hand, we seek to find a different counterterm that permits a perturbation expansion without any divergences. Superficially, it appears like we are seeking a `needle in a hay stack'; happily, in our case, we are able to find the `needle'!

   Although the formal
  functional integral (\ref{e2}) is essentially undefined, it can be given meaning by first introducing a lattice regularization in which the spacetime continuum is replaced with a periodic, hypercubic lattice with lattice spacing $a>0$ and with  $L<\infty$ sites along each axis. The sites themselves are labeled by multi-integers $k=\{k_0,k_1,\ldots,k_s\}\in{\mathbb Z}^n$, $n=s+1$, and $h_k$ and $\p_k$ denote field values at the $k$th lattice site; in particular, $k_0$ is designated as the Euclidean time variable. This regularization results in the $L^n$-dimensional integral
  \bn &&\hskip-2em S_{latt}(h)\equiv M\int e^{\t (1/\hbar)Z^{-{1/2}}\Sigma_k h_k\s\p_k\,a^n}\no\\
  &&\hskip3em\times e^{\t-(1/2\hbar)\s[\Sigma_{k,\s k^*}(\p_{k^*}-\p_k)^2\,a^{n-2}+\s m_0^2\s\Sigma_k\p_k^2\,a^n]}\no\\
  &&\hskip3em \times\,e^{\t-(1/\hbar)\s g_0\Sigma_k\s\p_k^4\,a^n-(1/2\s\hbar)\s\Sigma_k F_k(\p,\hbar)\,a^n}\;\Pi_k\s d\p_k   \no\\
    &&\hskip1.5em \equiv  \<\s e^{\t (1/\hbar)Z^{-{1/2}}\Sigma_k h_k\s\p_k\,a^n}\s\>\;. \label{e4} \en
  Here we have introduced the field-strength renormalization constant $Z$. Each of the factors $Z$, $m_0^2$, and $g_0$ are
  treated as bare parameters implicitly dependent on the lattice spacing $a$, ${k^*}$ denotes one of the $n$ nearest neighbors in the positive direction from the site $k$, and $M$ is chosen so that $S_{latt}(0)=1$.
  The counterterm $F_k(\p,\hbar)$ also implicitly depends on $a$, and
  the notation $F_k(\p,\hbar)$ means that the formal, locally generated counterterm $F(\p,\hbar)$ may, when lattice regularized,  depend on finitely-many field values located within a small, finite region of the lattice around the site $k$. Noting that a primed sum $\Sigma'_k$, below and elsewhere, denotes a sum over a single spatial slice all sites of which have the same Euclidean time $k_0$, we are led to choose the counterterm
  \bn \Sigma_k\s F_k(\p,\hbar)\s\s a^n\equiv \Sigma_{k_0}\s a\,[\hbar^2\Sigma'_k{\cal F}_k(\p)\,a^s]\en  with
                \bn {\cal F}_k(\p)\hskip-1.1em&& \equiv\quarter\s(1-2ba^s)^2\s
          a^{-2s}\s\bigg({\ts\sum'_{\s t}}\s\frac{\t
  J_{t,\s k}\s \p_k}{\t[\Sigma'_m\s
  J_{t,\s m}\s\p_m^2]}\bigg)^2\no\\
  &&\hskip2em-\half\s(1-2ba^s)
  \s a^{-2s}\s{\ts\sum'_{\s t}}\s\frac{J_{t,\s k}}{[\Sigma'_m\s
  J_{t,\s m}\s\p^2_m]} \no\\
  &&\hskip2em+(1-2ba^s)
  \s a^{-2s}\s{\ts\sum'_{\s t}}\s\frac{J_{t,\s k}^2\s\p_k^2}{[\Sigma'_m\s
  J_{t,\s m}\s\p^2_m]^2}\;. \label{eF} \en
  Here $b$ is a fixed, positive parameter with dimensions (Length)$^{-s}$ and $J_{k,l}=1/(2s+1)$ for $l=k$ and for $l$ one of the $2s$ spatially nearest neighbors to $k$; otherwise, $J_{k,l}=0$.
  Although ${\cal F}_k(\p)$ does not depend only on $\p_k$, it nevertheless becomes a local potential
  in the formal continuum limit.

  At first sight, it seems impossible that such a counterterm could be successful since it is strong for {\it small} field values while the usual divergence difficulties are due to {\it large} field values. It has even been suggested that such a term would vanish when the renormalization group is invoked. If the reader entertains similar thoughts, then this article is addressed to you because {\it this counterterm does the job!}

 While much of our discussion is straightforward, our discovery of the special counterterm will be  indirect.
   To begin, we observe that if $\<\s[\Sigma_k\s g_0\s\p^4_k\s\s a^n]^p\s\>$ denotes a {\it spacetime} lattice expectation value and 
      $\<\s[\S_{k}\s g_0\s\p^4_{k}\s\s a^s]^p\s\>$
  denotes a {\it spatial} lattice expectation value at a fixed Euclidean time $k_0$, then
  \bn &&\hskip-1.6em|\s\<\s[\Sigma_k\s g_0\s \p^4_k\s\s a^n\s]^p\s\>\s|=\Sigma_{k_{0_{(1)}}}\cdots\Sigma_{k_{0_{(p)}}}\s a^p\,|\s\<\{\,[\S_{k_{(1)}} g_0\s\p^4_{k_{(1)}}\s\s a^s]\cdots[\S_{k_{(p)}} g_0\s\p^4_{(p)}\s\s a^s]\,\}\s\>\s| \no\\
      && \hskip1.5em\le \,\Sigma_{k_{0_{(1)}}}\cdots\Sigma_{k_{0_{(p)}}}\s a^p\s\,|\,\<[\S_{k_{(1)}} g_0\s\p^4_{k_{(1)}}\s\s a^s]^p\>\cdots \<[\S_{k_{(p)}} g_0\s\p^4_{k_{(p)}}\s\s a^s]^p\>\,|^{1/p} \;, \label{hol}\en
      based on H\"older inequalities. It follows that if the {\it spatial} expectation value is finite then the {\it spacetime} expectation value is finite; clearly, if the spatial term diverges, then the spacetime term also diverges. Next, we recall that the ground state of a system determines the Hamiltonian operator, which in turn determines the lattice action including the relevant counterterm.
    Thus, in order to further the search for a new counterterm, our initial focus will be to examine the ground-state wave function.

     \subsection*{Determining the proper counterterm}
     We now return to the lattice-regularized functional integral (\ref{e4}). In order for
     this mathematical expression to be physically relevant following a Wick rotation to real time, we impose the requirement of {\it reflection positivity}
     \cite{gj}, which is assured if the counterterm satisfies
       \bn \Sigma_k\s F_k(\p,\hbar)\equiv\Sigma_{k_0}\s \{\Sigma'_k { F}_k(\p,\hbar)\} \;,\en
       where the sum over the Euclidean time lattice is outside the braces. Thus each element $\{\Sigma'_k { F}_k(\p,\hbar)\}$ of the temporal sum involves fields all of which have the same temporal value $k_0$, but they may involve several other fields at nearby sites to $k$ in spatial directions only.

       The lattice formulation (\ref{e4}) is a functional representation of the operator expression
         \bn S_{latt}(h)=N\s{\tr}\{\s \mathbb{T}\,e^{\t (1/\hbar)\Sigma_{k_0}\s[\s Z^{-1/2}\s\S_k \s h_k\s\hat{\p}_k\s\s a^n-\s \H\s\s a]}\}\;,\en
         where $N=\{{\tr}[e^{-(La)\s\H/\hbar}]\}^{-1}$ is a normalization factor, $\mathbb{T}$ signifies time ordering, $\hat{\p}_k$ is the field operator, and $\H$ denotes the Hamiltonian operator. When the function $h$ is limited to a single spatial slice, e.g., $h_k\equiv a^{-1}\s\delta_{k_0,0}\,f_k$, this expression reduces to
            \bn S'_{latt}(f)=N\s{\tr}\s\{\s e^{\t -(1/\hbar)(L-1)\H\s\s a}\, e^{\t (1/\hbar)\s[\s Z^{-1/2}\s\S_k\s f_k\s\hat{\p}_k\s\s a^s-\H\s\s a\s]}\}\en
             which, for large $La$, is realized by
       \bn S'_{latt}(f)=\int e^{\t(1/\hbar)\s Z^{-1/2}\s\Sigma'_k f_k\p_k\s\s a^s}\,\Psi_0(\p)^2\,\Pi'_kd\p_k\;, \label{e10}\en
       where $\Pi'_k$ denotes a product over a single spatial slice at fixed $k_0$,
       $\Psi_0(\p)$ denotes the unique, normalized ground-state wave function of the Hamiltonian operator $\H$ for the problem, expressed in the Schr\"odinger representation, with
       the property that $\H\Psi_0(\p)=0$.  After a change of integration variables from the $N'\equiv L^s$ variables $\{\p_k\}$ to  `hyperspherical coordinates' $\{\k, \eta_k\}$,  defined by $\p_k\equiv \k\s\eta_k$, $\k^2\equiv \Sigma'_k\p_k^2$, $1\equiv\Sigma'_k\eta_k^2$, where $0\le\k<\infty$, and $-1\le\eta_k\le1$, Eq.~(\ref{e10})
       becomes
         \bn S'_{latt}(h)\hskip-1.1em&&=\int e^{\t(1/\hbar)\s Z^{-1/2}\s \k\Sigma'_kh_k\s\eta_k\,a^s}\,\Psi_0(\k\s\eta)^2\,\k^{N'-1}\s d\k
         \s2\s\delta(1-\Sigma'_k\eta_k^2)\,\Pi'_k\s d\eta_k.\no\\  \label{e99}\en

         We define the continuum limit as $a\ra0$, $L\ra\infty$, with $La$ large and fixed.  In the continuum limit $N'=L^s\ra\infty$ and divergences will generally
         arise  in a perturbation expansion, because, when parameters like $m_0$ or $g_0$, are changed, the measures in (\ref{e99})
         become, in the continuum limit, {\it mutually singular} due to the overwhelming influence of $N'$ in the measure factor
         $\k^{N'-1}$. However, we can avoid that conclusion provided the ground-state distribution $\Psi_0(\p)^2$ contains a factor that serves to `mash the measure', i.e., effectively change $\k^{N'-1}$ to $\k^{R-1}$, where $R>0$ is fixed and finite.
         Specifically, we want the ground-state wave function to have the form
           \bn \hskip-1em\Psi_0(\p)\hskip-1em&&= ~^{``} M'\,e^{\t-U(\p,\hbar,a)/2}\,\k^{-(N'-R)/2}\s^{\,"}\no\\
           &&=M' e^{\t-U(\k\eta,\hbar,a)/2}\,\k^{-(N'-R)/2}
           \,\Pi'_k[\Sigma'_l\s J_{k,l}\s\eta^2_l]^{-(1-R/N')/4}\no\\
               &&=M' e^{\t-U(\p,\hbar,a)/2}\,\Pi'_k[\Sigma'_l\s J_{k,l}\s\p^2_l]^{-(1-R/N')/4}.\en
               The first line (in quotes) indicates the qualitative $\k$-behavior that will effectively mash the measure, while the second and third lines illustrate a specific functional dependence on field variables that leads to the
               desired factor. Here we choose the constant coefficients $J_{k,l}\equiv1/(2s+1)$ for $l=k$ and for $l$ equal to each of the $2s$ spatially nearest neighbors to the site $k$; otherwise, $J_{k,l}\equiv0$. Thus, $\Sigma'_lJ_{k,l}\s\p_l^2$ provides a
               {\it spatial average of field-squared values at and nearby the site $k$}. As part of the ground-state distribution, this factor is dominant for {\it small-field} values, and its form is no less fundamental than the rest of the ground-state distribution that is determined by the gradient, mass, and interaction terms that fix the {\it large-field} behavior. The factor $R/N'$ appears in the local expression of the small-field factor, and on physical grounds that quotient should not depend on the number of lattice sites in a spatial slice nor on the specific parameters mentioned above that define the model. Therefore, we can assume that $R\propto N'$, and so we set
                              \bn   R\equiv 2\s b\s a^s\s N'\;, \en
               where $b>0$ is a fixed factor with dimensions (Length)$^{-s}$ to make $R$ dimensionless. Even though the ground-state distribution diverges when certain of the $\eta_k$-factors  are simultaneously zero, these are all integrable singularities since when any subset of the $\{\eta_k\}$ variables are zero, there are always fewer zero factors arising from the singularities thanks to the local averaging procedure; indeed, this very fact has motivated the averaging procedure. Thus, the benefits of measure mashing are:
If a perturbation expansion of a certain term in the exponent $U(\p,\hbar,a)$ is made after $\k^{N'-1}$ is changed to $\k^{R-1}$, then the various measures are {\it equivalent} and the terms in the perturbation expansion are all finite. Since the perturbation-expansion spatial terms are finite, then, according to the inequality (\ref{hol}), so too are the corresponding spacetime terms.

               To obtain the required functional form of the ground-state wave function for small-field values, we choose our counterterm $F_k(\p,\hbar)$ to build that feature into the Hamiltonian. In particular, the counterterm is a specific potential term of the form
               $\half \Sigma'_k\s{ F}_k(\p,\hbar)\,a^s\equiv \half\hbar^2\Sigma'_k{\cal F}_k(\p)\,a^s$
               where, with $T(\p)\equiv \Pi'_r[\Sigma'_l J_{r,l}\s\p_l^2]^{-(1-2ba^s)/4}$,
                \bn {\cal F}_k(\p)\hskip-1.1em&& \equiv\frac{ a^{-2s}}{ T(\p)}\frac{\d^2 T(\p)}{\d\p_k^2}\en
               which leads to the expression in (\ref{eF}).

  More generally, the full, nonlinear Hamiltonian operator including the desired counterterm is defined as
  \bn \H\hskip-1.1em&&=-\half\hbar^2\s a^{-2s}{\t\sum}'_k\,\frac{\d^2}{\d\p_k^2}\,a^s+\half{\t\sum}'_{k,k^*}(\p_{k^*}-\p_k)^2\s a^{s-2} \no\\
    &&\hskip2em+\half\s m_0^2{\t\sum}'_k\p_k^2\,a^s+
    g_0{\t\sum}'_k\p_k^4\,a^s
    +\half\hbar^2{\t\sum}'_k{\cal F}_k(\p)\,a^s-E_0\;. \label{eH}\en
    Indeed, this latter equation implicitly determines the ground-state wave function $\Psi_0(\p)\s$!

    The author has used several different arguments previously  \cite{IOP} to suggest this alternative model for nontrivial scalar field quantization. The present article argues, simply, that an unconventional ${\cal O}(\hbar)$ counterterm added to the naive Hamiltonian is all that is needed to achieve viable results. After all, conventional quantization assumes---even when using the favored Cartesian coordinates---that the proper Hamiltonian operator is open to various quantum corrections that only experiment or mathematical consistency can resolve. In a nutshell, that is enough justification for our proposal.

    It is important to observe that no normal ordering applies to the terms in the Hamiltonian. Instead,
    local field operator products are determined by an operator product expansion \cite{qft}. In addition, note that the counterterm $\hbar^2\Sigma'_k{\cal F}_k(\p)$ does {\it not} depend on any parameters of the model and
    specifically not on $g_0$. This is because the counterterm is really a counterterm
    for the {\it kinetic energy}. This fact follows because not only is $\H\s\Psi_0(\p)=0$, but then $\H^q\s\Psi_0(\p)=0$ for all integers $q\ge2$. While $[\Sigma'_k\d^2/\d\p_k^2]\s\Psi_0(\p)$ may be a square-integrable function, the expression  $[\Sigma'_k\d^{2}/\d\p_k^{2}]^q\s\Psi_0(\p)$ will surely not be square integrable for suitably large $q$. To ensure that $\Psi_0(\p)$
    is in the domain of $\H^q$, for all $q$, the derivative term and the counterterm must be considered together to satisfy domain requirements, hence
    our claim that the counterterm should be considered as a `renormalization' of the kinetic energy.

    We recall the concept of a pseudofree theory \cite{book}, which is well illustrated by an anharmonic oscillator model with the classical action functional $A_g\equiv A_0+g\s A_I=\tint_0^T\{\half[\s{\dot x}(t)^2-x(t)^2]-g\s x(t)^{-4}\s\}\,dt$. It is clear that the domain $D(A_g)$ of allowed paths, $\{x(t)\}_0^T$, satisfies $D(A_g)=D(A_0)\cap D(A_I)$, and it follows that $\lim_{g\ra0}\s D(A_g)\equiv D(A'_0)\subset D(A_0)$.  In brief, the interacting model is {\it not} continuously connected to its own free theory $(A_0)$, but to a pseudofree theory $(A'_0)$, which has the same formal action functional as the free theory but a strictly smaller domain. This behavior persists in the quantum theory since, when $g\ra0$, the propagator for the interacting model does not pass continuously to the free quantum propagator but to a pseudofree quantum propagator with different eigenfunctions and eigenvalues. We now argue that a similar behavior may be seen in our scalar field models.

    Since the counterterm does not depend on the coupling constant, it follows that the counterterm remains even when
    $g_0\ra0$, which means that the interacting quantum field theory does {\it not} pass to the usual free quantum field theory as $g_0\ra0$, but instead it passes to a pseudofree quantum field theory. To help establish this point, let us first consider the classical (Euclidean)
        action functional (\ref{e1}). Regarding the separate components of that expression,
        and assuming $m_0>0$, a multiplicative inequality \cite{russ,book} states that
          \bn \{\tint \p(x)^4\,d^n\!x\}^{1/2}\le{\tilde C}\tint[(\nabla\p)(x)^2+m_0^2\p(x)^2]\,d^n\!x\label{e333}\;,\en
          where ${\tilde C}= (4/3)\s [\s m_0^{(n-4)/2}\s]$ when $n\le4$ (the renormalizable cases), and ${\tilde C}=\infty$ when  $n\ge5$ (the nonrenormalizable cases), which means, in the latter case, there are
          fields $\p(x)$ for which the integral on the left side of the inequality diverges
          while the integral on the right side is finite; for example $\p_{sing}(x)=|x|^{-\a}\s e^{-x^2}$, for $n/4\le\a< (n-2)/2$.
          In other words,  for $\p^4_n$ models in particular, there are different pseudofree and free classical theories when $n\ge5$. Thus, for $n\ge5$, it is reasonable to assume that the pseudofree quantum field theory is different from the free quantum field theory.

          It is significant that a similar inequality
             \bn \{\tint \p(x)^p\,d^n\!x\s\}^{2/p}\le {\tilde C}' \tint[(\nabla\p)(x)^2+m_0^2\p(x)^2]\,d^n\!x\label{e333p}  \;,\en
             where ${\tilde C}'= (4/3)\s [\s m_0^{[(p-2)n/p-2]}\s]$
          holds when  $p\le 2n/(n-2)$ (the renormalizable cases), and ${\tilde C}' =\infty$ when  $p> 2n/(n-2)$ (the nonrenormalizable cases) \cite{book}. Since our choice of counterterm did not  rely on the power four  of the interaction term, it follows that the very same counterterm applies to any interaction of the form $\tint \p(x)^p\,\dn$, as well as a sum of such terms, provided that the potential has a lower bound.

          Consequently,  the quantum models developed in this article with the unconventional counterterm provide
   viable candidates for those quantum theories normally classified as nonrenormalizable, and they do so
   in such a manner that in a perturbation analysis divergences do not arise because all the underlying measures are equivalent and not mutually singular. (A discussion of the divergence-free properties from a perturbation point of view appears in \cite{IOP}, an analysis that also determines the dependence
   of $Z$, $m_0^2$, and $g_0$ on the parameters $a$ and $N'$.) Since the unconventional counterterm conveys good properties to the nonrenormalizable models, it is natural to extend such good behavior to the traditionally renormalizable models by using the unconventional counterterm for them as well.
   {\it Thus, we are led to adopt the lattice regularized Hamiltonian $\H$ (\ref{eH}), including the counterterm, for all spacetime dimensions $n\ge2$.} Of course, the conventional quantizations for $\p^4_n$, $n=2,3$, are still valid for suitable physical applications of their own.

   Additionally, the lattice Hamiltonian (\ref{eH}) also determines the lattice Euclidean action including the unconventional counterterm, which is then given by
     \bn I=\half \Sigma_{k,k^*}(\p_{k^*}-\p_k)^2\,a^{n-2}+\half m_0^2\Sigma_k\p_k^2\s a^n
     +g_0\Sigma_k\p_k^4
     \s\s a^n+\half\Sigma_k \hbar^2{\cal F}_k(\p)\s\s a^n, \label{eL}\no\\ \en
     where in this expression $k^*$ refers to all $n$ nearest neighbors of the site
     $k$ in the positive direction. Although the lattice form of the counterterm involves averages over field-squared values
     in nearby spatial regions of the central site, it follows that the continuum limit of the counterterm
     is local in nature, as noted previously. It is noteworthy that preliminary Monte Carlo studies
     based on the lattice action (\ref{eL}) support a nontrivial
     behavior of the $\p_4^4$ model exhibiting a positive renormalized coupling constant \cite{stank}.

\subsection*{Comments}   
  There already exist well-defined, conventional results for $\p^4_2$ and $\p^4_3$, and, yes, for different applications, we are proposing alternative quantizations for these models. The unusual counterterm (\ref{eF}) depends on the number of spacial dimensions, but it does not depend on the power $p$ of the interaction $\p^p_n$. Thus the counterterm applies to `mixed models' like $g_{0}\s\p^4_3+g'_{0}\s\p^8_3$, $g''_0\s\p^4_5+g'''_0\s\p^8_5+g''''_0\s\p^{138}_5$, etc., which exhibit reasonable properties when the coupling constants are turned on and off, in different orders, again and again.
  Unlike the conventional approach, there are no divergences in perturbative expansions about the pseudofree model in the new formulation after operator product expansions are introduced for the local operators. Although all of the terms in a perturbation expansion are finite, they cannot be computed by using Feynman diagrams because the validity of such diagrams relies on the functional integral for the free model having a Gaussian distribution, which is not the case for the pseudofree model.
For the important example of $\p^4_4$, traditional methods lead to a renormalizable theory, yet nonperturbative methods support a trivial (= free) behavior. On the other hand, our procedures for $\p^4_4$ are expected to be divergence free and nontrivial.

\section*{Dedication}
It is a pleasure to dedicate this article to Prof.~Andrey Slavnov on the celebration of his 75th birthday. Andrey remains an outstanding scientist as well as a long-time friend and colleague of the first rank. It is an honor to wish him continued good health and creative work in the future.

\end{document}